# Evolution of phase morphology in dispersed clay systems under the microwave irradiation

*(Evolução da morfologia de fases em sistemas dispersos de argila sob irradiação de micro-ondas)*

*A. G. Chetverikova*[1]\*, M. M. Filyak*[1]*, O. N. Kanygina*[1]
[1]*Orenburg State University, Orenburg, Russia*

**Abstract**

The results of a study of the microwave emission effect (power 700 W, frequency 2.45 GHz) on the structural changes in natural clay particles with effective diameters D≤630 μm are presented. The influence of the irradiation time (10 and 20 min) and the environment in the microwave chamber (atmospheric air and air saturated with water vapor) on the structural changes occurring in the particles were traced. During the first 10 min, capillary water was completely removed and agglomeration was carried out by attaching single dispersed particles (diffusion limited aggregation model). At the second stage (10-20 min), the already formed agglomerates grew (cluster-cluster aggregation model). A complex of independent optical-physical methods was used to analyze weak structural changes, including X-ray diffraction analysis, colorimetry and wavelet analysis. This approach increased the information content and reliability of measurements, quantitatively characterizing the structural responses in disperse clay systems. In the air, the removal of capillary water was accompanied by agglomerations of particles and polymorphic transformations of oxides: montmorillonite was completely decomposed and amorphous phases, in particular $CaCO_3$, crystallized. The composition of the environment in the microwave chamber affected the type of phase transformations in iron compounds: iron-aluminum silicate ($Fe_2Al_4Si_5O_{18}$) was formed in the air; magnetite $Fe.Fe_2O_4$ ($Fe_3O_4$) appeared in the water vapor medium. The carried-out studies with the developed set of experimental methods indicated the possibility of regulating the processes of structure formation in dispersed clay systems by optimizing the regimes of exposure to microwave radiation.

*Resumo*

*Os resultados do estudo do efeito de emissão de micro-ondas (potência 700 W, frequência de 2,45 GHz) nas mudanças estruturais em partículas de argila natural com diâmetros efetivos D≤630 μm são apresentados. A influência do tempo de irradiação (10 e 20 min) e o ambiente na câmara de micro-ondas (ar atmosférico e ar saturado com vapor de água) nas mudanças estruturais que ocorrem nas partículas foi rastreada. Durante os primeiros 10 min, a água capilar foi completamente removida e a aglomeração ocorreu pela junção de partículas dispersas individuais (modelo de agregação limitada por difusão). No segundo estágio (10-20 min), os aglomerados já formados (modelo de agregação cluster-cluster) cresceram. Um complexo de métodos óptico-físicos independentes foi utilizado para analisar mudanças estruturais fracas, incluindo análise de difração de raios X, colorimetria e análise de wavelet. Essa abordagem aumentou o conteúdo da informação e a confiabilidade das medidas, caracterizando quantitativamente as respostas estruturais nos sistemas dispersos de argila. No ar, a remoção de água capilar foi acompanhada por aglomerações de partículas e transformações polimórficas de óxidos: a montmorilonita foi completamente decomposta e as fases amorfas, em particular, $CaCO_3$, cristalizaram. A composição do ambiente na câmara de micro-ondas afetou o tipo de transformação de fase em compostos de ferro: silicato de ferro-alumínio ($Fe_2Al_4Si_5O_{18}$) foi formada no ar; magnetita $Fe.Fe_2O_4$ ($Fe_3O_4$), apareceu no meio de vapor de água. Os estudos realizados com o conjunto desenvolvido de métodos experimentais indicaram a possibilidade de regular os processos de formação de estruturas em sistemas dispersos de argila, otimizando os regimes de exposição à radiação de micro-ondas.*
*Palavras-chave: morfologia de fase, sistema disperso, aglomeração, argila de montmorilonita, radiação de micro-ondas, resposta estrutural.*

# INTRODUCTION

Investigation of the interrelation composition-technology-structure-properties in dispersed systems, consisting of clay minerals particles, requires an increase in the level of measurements informativeness. This problem is especially relevant while studying the weak structural responses of a system to the microwave (MW) effect. In recent years, microwave heating has been widely employed in the sintering and joining of ceramics. Microwave sintering has characteristics such as low-temperature and short time heating. The use of microwave processing methods significantly reduces energy consumption, particularly in high-temperature processes. [1]. However, the advantages of using microwave energy in high-temperature processes are not limited to energy savings. Generally, it produces better electrical and mechanical properties than the conventional ones. Authors [2] have studied the benefits of the microwave technology for sintering on montmorillonite with Al-polycations. They found that the microwave technology gave a unique possibility of influencing the microstructures and physical properties of the ceramics. Recent publications [3, 4] show ambiguity in interpreting the results of experimental studies of structural transformations in aluminosilicates subjected to MW effects due to weak structural responses. As before, one can not speak of a general understanding of the modification mechanisms, the formation of the secondary phase morphologies in dispersed clay systems, despite the use of a number of modern and classical experimental methods.

*\*KR-727@mail.ru*

Taking into account the complex structure of natural aluminosilicate materials, we can say that traditional methods of quantitative description of structure formation processes fit poorly. Our experience shows that for the registration of weak structural responses of dispersed clay systems, a set of experimental methods is required that simultaneously detect the generation of the secondary phase morphologies at the millimeter, micron, and submicron levels [5]. Along with optical microscopy, which provides images of dispersed samples surfaces, modern computer processing of these images is required, with the help of methods of colorimetric gradation, methods of multifractal and wavelet analyses. To study the morphology of powder materials, information is needed on the change in the color parameters of their surface, the periodicity of the structural elements, the nature of their distribution in the volume, and scale invariance. The algorithm for studying the processes of structural transformations in a dispersed clay system exposed to MW radiation, in our opinion, should be as follows: i) analysis of the morphology of the dispersed system at the meso-level by processing optical images using colorimetric gradation methods, fractal and wavelet analyses; ii) registration of the formation of secondary phase morphologies at the micron and submicron levels by X-ray patterns analysis. The method of colorimetric gradation is very effective in computer processing of optical images of clay powders surfaces [6]. In modern colorimetry, several color models of different target orientation are operating [7], however, for the analysis of structural changes in aluminosilicates, the XYZ and L*a*b* models are the most convenient [8]. Recently, information on the periodicity and nature of the distribution of structural elements in the material was obtained using the wavelet analysis. This method is used to process signals depending on the time or spatial coordinate. While analyzing the surface of a solid body in optical systems for recording and processing information, a spatial coordinate is used [9].

The objective of this paper is to study the weak structural changes occurring in dispersed clay systems under the action of microwave radiation with the optical-physical methods. The complex of independent methods, including X-ray phase analysis, colorimetry and wavelet analysis, makes it possible to increase the informativity and reliability of measurements, quantify self-organization processes and structural responses in dispersed clay systems by correlating experimental data.

**EXPERIMENTAL**

The object of investigation was powder of native clay, containing montmorillonite; the chemical composition is given in Table I [10]. A batch of clay power used in our experiments had mineral particles with effective diameters D≤630 μm. All powder samples were filled into a cylindrical cuvette with a layer thickness of 10 mm and irradiated by microwave fields (frequency f=2.45 GHz, power 700 W) in standard condition. MW field was produced by the magnetron. Sample heating was due to microwave field absorption by electrically conductive regions of clay particles. As far as microwave field absorption is inhomogeneous in the volume [11] all the samples were placed on a rotating plate for averaging of MW heating conditions. The temperature of samples was measured by chromel-alumel thermocouple. For temperature measurements, the thermocouple was put down inside samples to a depth of 5 mm just after switching off microwave irradiation. The time delay between turning off the field and the temperature measurements did not exceed 20 s. Microwave treatment was produced in two ways: 1) in the air for 10 and 20 min (lots B1 and B2), while the temperature of the powder samples reached 200 °C; 2) in a water vapor medium for 20 min, at a temperature of powder samples of about 400 °C (batch of sample C).

Structural changes at the meso-level were estimated from optical images using colorimetric gradation and fractal analysis. Surface images were obtained by using a metallographic type microscope with a digital high-resolution eyepiece in the magnification range from 10 to 1000 times. To assess the color difference, the colorimetric systems XYZ and CIE (Commission Internationale de l'Éclairage) L*a*b* were used, where x and y are coordinates in the color triangle of the ICI (International Commission on Illumination). In the CIE model L*a*b*, L* denotes lightness, a* is the red/green component, and b* is the yellow/blue component. The color difference was determined by

Table I - Chemical composition (wt%) of native clay.
*[Tabela I - Composição química (% em massa) da argila nativa.]*

| $SiO_2$ | $Fe_2O_3$ | $TiO_2$ | $Al_2O_3$ | CaO | MgO | $Na_2O$ | $K_2O$ | LOI | Σ |
|---|---|---|---|---|---|---|---|---|---|
| 55.9 | 9.51 | 0.86 | 18.63 | 0.72 | 2.05 | 1.9 | 3.24 | 7.08 | 99.89 |

*LOI - loss on ignition.*

the formula $\Delta E^*_{ab}=[(\Delta L^*)^2+(\Delta a^*)^2+(\Delta b^*)^2]^{1/2}$. Fractal analysis of the morphology of dispersed samples was carried out using the ImageJ program and an additional FracLac 2.5 module [12]. Fractal dimensions were estimated by the grid method, in which the image of the object was divided into a set of cells of specified sizes. The image scanning of dispersed systems was carried out in several cycles, while on each subsequent cycle the dimensions of the grid cells were increased. The fractal dimension of the analyzed object was calculated by the formula [13]:

$$D_s = \lim_{\varepsilon \to 0} \frac{\ln M(\varepsilon)}{\ln \frac{1}{\varepsilon}} \quad (A)$$

where $M(\varepsilon)$ is the minimum number of cells with side $\varepsilon$ required to cover all elements of the image. A significant parameter in the description of stochastic fractal structures is lacunarity [14], a measure of the inhomogeneity of the space filling by an object. For its calculation the formula [14] was used:

$$\Lambda = \left(\frac{\sigma}{\mu}\right)^2 \quad (B)$$

where $\sigma$ is the standard deviation of the unit image elements of the sample in grid cells of a given size $\varepsilon$; $\mu$ is the average value of the unit image elements of a sample in cells of a specified size $\varepsilon$. Lacunarity, as the fractal dimension, is determined from the slope of the regression line in the coordinates $\ln\varepsilon=f(\ln\Lambda)$. As it was shown earlier [15], variations in the values of fractal dimensions for powders, consisting of clay minerals particles, make it possible to control changes in the specific surface area of systems, and to establish the formation of self-ordered structures. The wavelet transform of an arbitrary function $z(x)$ was defined as an integral transformation of the form:

$$W(m,n)=\frac{1}{\sqrt{m}}\int_{-\infty}^{+\infty} z(x)\psi\left(\frac{x-n}{m}\right)dx \quad (C)$$

The result of the wavelet transform is the spectrogram $W(m,n)$ - a function of two variables: the scale parameter of the wavelet - m, and the wavelet shift parameter - n. The wavelet analysis is called a 'mathematical microscope'; it preserves resolution at different scales, reveals the internal structure of a substantially heterogeneous object and allows one to study its local features. The shift parameter n sets the focusing point of the microscope, the scale parameter m is an increase; the choice of the basis wavelet determines the optical qualities of the microscope [16]. By the comparative analysis of different wavelet bases, it was established that the most accurate and complete identification of the features of the profile of physical surfaces is the Morlet wavelet, described by the expression:

$$\psi(t) = e^{i\omega_0 t}e^{-t^2/2} \quad (D)$$

where $\omega_0$ is the center frequency of the wavelet. X-ray diffraction (XRD) patterns for the phase analysis of the phase compositions of clay systems were obtained with a MD10 mini diffractometer in monochromatized copper radiation with scanning rates of 1/8 and 1 °/min in the range of angles from 10° to 120° to the scale of 2θ (±0.02°) with recording of the results in digital format; crystalline phases were identified with the LookPDF program using the International Center for Diffraction Data (ICDD) files.

**RESULTS AND DISCUSSION**

*Influence of the time of MW radiation on structural changes in clay*

The temperature in the volume of powders treated in the MW chamber rose to 160-170 °C after 10 min of irradiation and up to 190-200 °C in 20 min; the loss of mass due to the evaporation of water was 3.5 to 5.5%, respectively. Under the effect of MW radiation on clay minerals particles, a structure was formed and it could be considered as a hierarchy that satisfied the property of self-similarity, that is, a fractal one. The agglomeration of particles, which depended on the time of the MW radiation, was developing. Fig. 1 shows the surface images of powders A, obtained before (Fig. 1a, batch A), and after MW processing for 10 min (Fig. 1b, batch B1) and 20 min (Fig. 1c, batch B2). In Fig. 1a fragmentary particles are visible. After 10 min of microwave irradiation, they were 'overgrown' with small particles (Fig. 1b). Agglomerates separated by large cracks appeared after irradiation for 20 min (Fig. 1c).

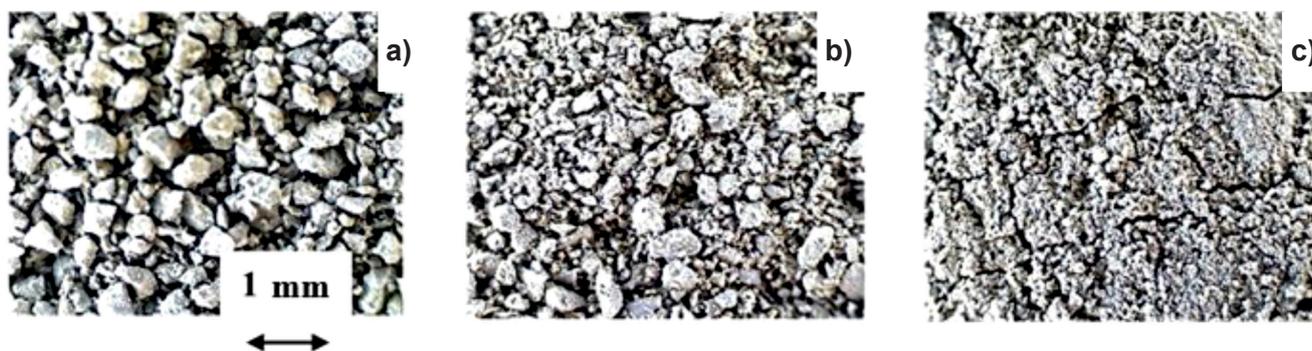

Figure 1: Optical images of the surfaces of a polydispersed system in the initial state (a), and after 10 min (b) and 20 min (c) of exposure to MW radiation.
*[Figura 1: Micrografias obtidas por microscopia óptica das superfícies de um sistema polidisperso no estado inicial (a) e após 10 min (b) e 20 min (c) de exposição à radiação de micro-ondas.]*

The agglomeration was indirectly confirmed by the methods of colorimetry: at close values of the color parameters R, G, B (grey color), the values of reflection indexes for the surfaces of powders A, B1 and B2 decreased as 1:0.9:0.7. This fact can be explained with the help of the results of the fractal analysis. The surfaces of all three sample batches had a developed terrain and an irregular structure. The $D_s$ values of the fractal dimensions exceed the topological one and did not change with increasing images, which confirmed the fractal nature of the mesostructure. It was established that the values of the fractal dimension $D_s$ increased from 2.826 to 2.852 with the increasing of MW exposure time (Fig. 2, line 2). The increase in the fractal dimension was associated with the agglomeration of powder particles under the effect of MW radiation. Due to the fractality of the structure of the samples, a local increase in the intensity of the electromagnetic field occurred, leading to the generation of ponderomotive forces proportional to the gradient of the square of the field amplitude $-\vec{\nabla}|\vec{E}|^2$. Under the effect of these forces, charged particles were pushed out of areas with high electromagnetic field strength into an area with a lower one [17]. There takes place spatial redistribution, self-organization of interacting particles - agglomeration.

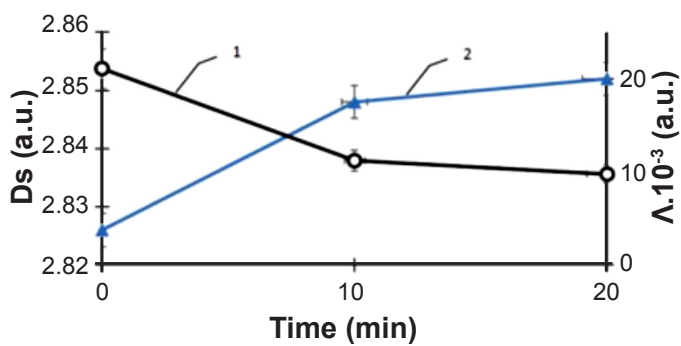

Figure 2: Dependences of fractal dimension Ds and lacunarity Λ of aluminosilicate particles in the initial state, and after 10 and 20 min of exposure to MW radiation: 1 - Λ, 2 - Ds.
[Figura 2: Dependências da dimensão fractal Ds e lacunaridade Λ de partículas de aluminossilicato no estado inicial e após 10 e 20 min de exposição à radiação de micro-ondas: 1 - Λ, 2 - Ds.]

The formation of agglomerates in the first 10 min can be described by the DLA (diffusion limited aggregation) model [14, 18] when the agglomerate grows due to the addition of dispersed single particles. Such agglomerates have a compact structure, which leads to densification of the sample and an increase in the fractal dimension. With a longer exposure, the already formed agglomerates are involved in the growth process, and another agglomeration model, the cluster-cluster aggregation (CCA) model, comes to the fore [14, 18]. At this scale level, the cluster agglomerate is formed by combining smaller agglomerates. New agglomerates have a more friable structure, which affects the decrease in the growth rate of fractal dimension. The formation of agglomerates at different stages according to DLA and CCA models is confirmed by Fig. 2, where the lacunarity of the samples Λ is shown as a function of the treatment time. It is known that for the emptier areas, the higher is the value of Λ. In the first 10 min of MW exposure, when the compaction elements were individual particles, lacunarity decreased. With the formation of loose agglomerates, the value of lacunarity practically did not change. The formation of agglomerates by the first and, especially, the second mechanism led to the formation of less reflective surfaces, and hence to a decrease in the values of the reflection indexes.

*Influence of the external environment on structural changes in clay*

Evolution of the morphology of dispersed particles after MW treatment in air and water vapor for 20 min was monitored by optical images of surfaces treated with colorimetry and wavelet analysis. The most sensitive to changes in color characteristics are the parameters L*, a*, b*. According to the program developed by the authors, the color parameters L*, a* and b* for three batches of samples were calculated (Table II). It is seen that the lightness L* decreased insignificantly in the air after microwave exposure, and 1.5 times in the presence of water vapor. Variations in the parameters a* and b* confirmed the differences in the evolution of structural transformations occurring in a dispersed clay system under microwave treatment in different media. The most reliable [8] indicator of the structural changes presence is the color difference $\Delta E^*_{ab}$. If for sample A we take $\Delta E^*_{ab}=0$, then for the sample of lot B $\Delta E^*_{ab}=4.90$. In sample C, the color difference was 6 times higher ($\Delta E^*_{ab} \approx 30$). According to the generally accepted ideas, the condition $\Delta E \geq 2$ indicates the occurring changes in phase compositions, phase morphology. In an atmosphere saturated with water vapor, structural changes were more significant. This conclusion was confirmed by the results of wavelet analysis.

Fig. 3 shows wavelet spectrograms carrying information on the distribution of mesostructural elements (particles) on the surface of powders. The abscissa is the wavelet shift parameter, the ordinate is the scale value of the basis wavelet in conventional units. At the bottom of the spectrogram (1-3) there are small-scale values that reflect small structural elements. The middle part of the wavelet spectrum (3-7) represents the distribution of larger elements; the upper part (7-11) gives a generalized picture of the surface. The correlation between the wavelet and the point profile in its vicinity is determined by the color of the corresponding region of the wavelet spectrogram [16]: white is the maximum and dark blue is the minimum of coincidences. The presence of repeating similar elements - blue arches with green borders - confirmed the self-similarity and fractal properties of surface morphology established earlier in the article [15]. For the initial sample (Fig. 3a) on 1-3 scales, the green areas corresponded to the content of fine particles. On the spectrograms of batches subjected to microwave radiation (Figs. 3b and 3c), up to 4-5 scales of the correlation scales between the surface profile and the wavelet were absent (dark blue areas), small particles adhered to the formation of agglomerates. The tops of the arches after the MW irradiation shifted in scale from 3.5 to 4.5 (Fig. 3a) to 6-7 (Figs. 3b and 3c). This effect is explained by the agglomeration mechanisms of the powder particles described above.

Table II - Values of the color parameters in the system L*a*b*.
[Tabela II - Valores dos parâmetros de cor no sistema L*a*b*.]

| Batch | L* | a* | b* | $\Delta E^*_{ab}$ |
|---|---|---|---|---|
| A | 60.3 | 3.27 | -1.4 | 0 |
| B2 | 56.1 | 5.52 | 0.01 | 4.90 |
| C | 41.1 | 3.36 | -4.64 | 29.97 |

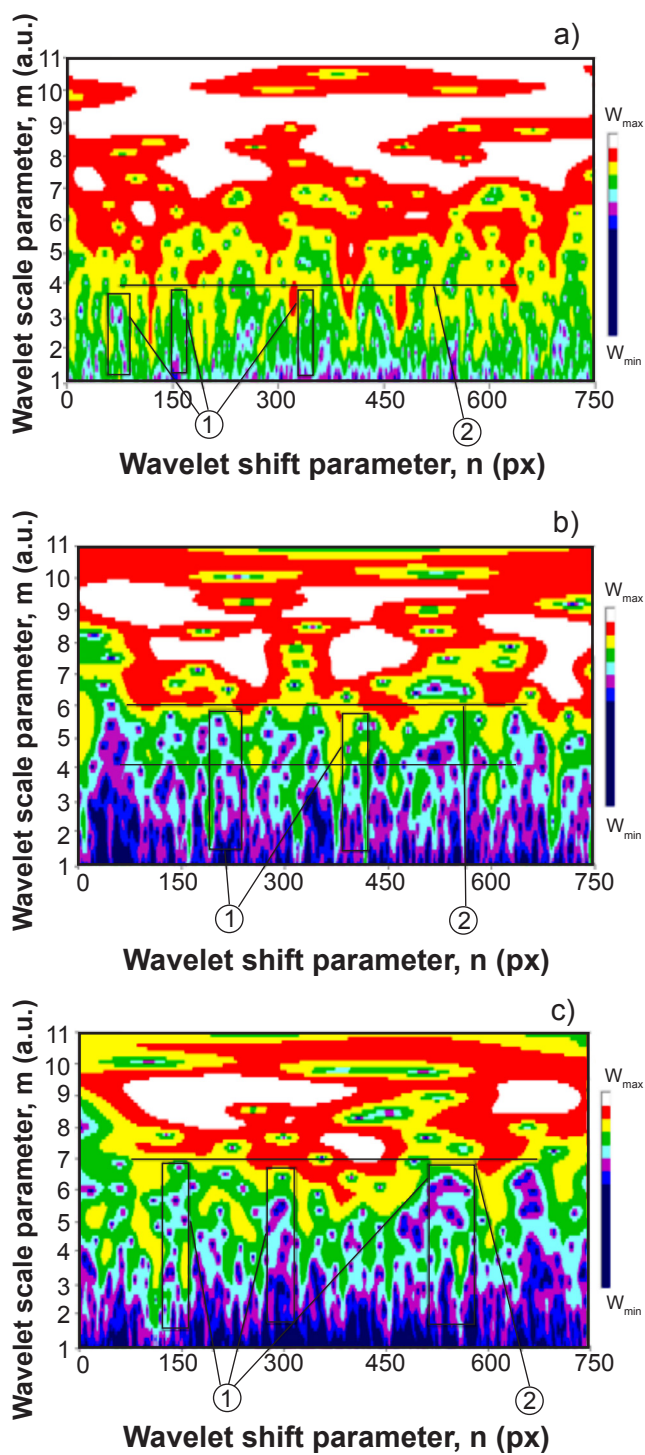

Figure 3: Wavelet spectrograms of dispersed systems: initial (a), and after 20 min of MW treatment in air (b), and in water vapor (c); 1 - arches, 2 - average level of apexes of arches.
*[Figura 3: Espectrogramas de wavelet de sistemas dispersos: inicial (a) e após 20 min de tratamento por micro-ondas no ar (b) e em vapor de água (c); 1 - arcos, 2 - nível médio de ápices de arcos.]*

Structural responses of the millimeter and micron levels to MW radiation are provided by submicron phase composition changes. X-ray diffraction analysis made it possible to establish the content of the main phases in complex multicomponent clay systems. The phase compositions of all three batches are presented in Fig. 4. Differences in the phase compositions again confirm the presence of weak structural changes occurring in the dispersed clay system under the effect of the microwave field, as well as the influence of the external environment on the nature of these changes. The total fraction of crystalline phases in batch B2 did not change, however, phase and polymorphic transformations took place. Instead of magnetite $Fe.Fe_2O_4$ ($Fe_3O_4$), $Fe_2O_3$ generated iron-aluminum silicate ($Fe_2Al_4Si_5O_{18}$). The content of montmorillonite decreased and the volume fraction of corundum ($Al_2O_3$) increased. The microwave effect on the dispersed system in a medium containing water vapor led to a 1.7-fold increase in the fraction of crystalline phases due to the burnout of organic matter and the partial crystallization of amorphous substances. Batch C contained 9 identifiable crystalline phases (instead of 6 in samples A and B2). Three new phases were observed: sillimanite, zeolite and calcite ($CaCO_3$); the volume fraction of magnetite and corundum increased, and montmorillonite decomposed.

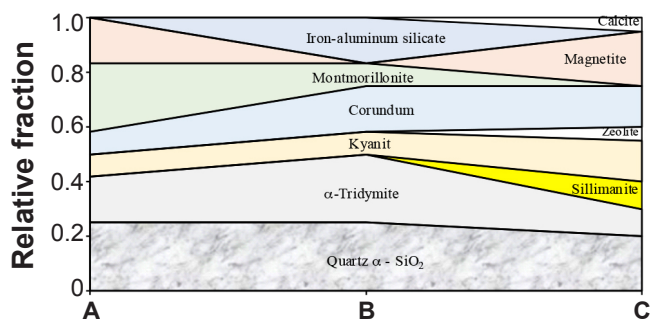

Figure 4: Phase compositions of the dispersed system before and after microwave exposure.
*[Figura 4: Composições de fase do sistema disperso antes e depois da exposição às micro-ondas.]*

## CONCLUSIONS

We report that the results of a study of microwave radiation effect (power 700 W, frequency 2.45 GHz) on the structural changes in natural clay particles with effective diameters D≤630 µm lead to the following conclusions. In a dispersed system, structural changes occur at the millimeter, micron and submicron levels. Based on the analysis of experimental data (computer processing of optical images of surfaces and diffractograms), we concluded that, together with the removal of capillary water, polymorphic transformations occur accompanied by particle agglomeration. The nature and intensity of structural responses depend on the exposure time and the environment in the microwave chamber. So, two-time intervals - 10 and 20 min, were considered. During the first 10 min, capillary water was completely removed and agglomeration was carried out by attaching single dispersed particles (diffusion limited aggregation model). At the second stage (10-20 min), already formed agglomerates grew (cluster-cluster aggregation model). The environment saturated with water vapor activated polymorphic transformations in the particles: montmorillonite was completely decomposed; amorphous phases, in particular $CaCO_3$, crystallized. There were new aluminosilicates - sillimanite ($Al_2O_3.SiO_2$) and zeolite ($Al_2O_3.54SiO_2$). The composition of the environment in the microwave chamber affected the type of phase transformations in iron compounds: iron-aluminum silicate ($Fe_2Al_4Si_5O_{18}$) was formed in air, magnetite $Fe.Fe_2O_4$ ($Fe_3O_4$) appeared in the water vapor medium. The developed set of studies significantly increased the level of informativeness of the experimental results and indicated the possibility of regulating the processes of structurization in dispersed clay systems by optimizing the microwave radiation regimes.

## ACKNOWLEDGEMENT

This work was performed at the support of the grant RFBR No. 17-42-560069 r_a "New optic-mathematical methods for analysis of structural variations of natural dispersed and nanostructured system".

## REFERENCES

[1] R.R. Menezes, P.M. Souto, R.H.G.A. Kiminami, in: Sintering of ceramics - new emerging techniques, A. Lakshmanan Ed., Intech, Rijeka (2012) 3.
[2] B. González, A.H. Pérez, R. Trujillano, A. Gil, M.A. Vicente, Materials **10**, 8 (2017) 886.
[3] I.A. Zhenzhurist, Glass Ceram. **72**, 7-8 (2015) 262.
[4] Y.V. Bykov, S.V. Egorov, A.G. Eremeev, V.V. Kholoptsev, I.V. Plotnikov, K.I. Rybakov, A.A. Sorokin, Materials **9**, 8 (2016) 684.
[5] V.M. Mel'nik, V.D. Rud', Y.A. Mel'nik, Powder Metall. Met. Ceram. **53**, 1-2 (2014) 107.
[6] A.G. Chetverikova, O.N. Kanygina, Meas. Tech. **59**, 11 (2016) 618.
[7] T.B. Gorshkova, Meas. Tech. **48**, 10 (2005) 1096.
[8] Y.T. Platov, D.A. Sorokin, R.A. Platova, Glass Ceram. **66**, 3-4 (2009) 125.
[9] V.L. Hilarov, V.E. Korsukov, V.N. Svetlov, P.N. Butenko, Phys. Solid State **46**, 10 (2004) 1868.
[10] M.M. Filyak, A.G. Chetverikova, O.N. Kanygina, L.S. Bagdasaryan, Condens. Matter. Interphases **18**, 4 (2016) 578.
[11] S.V. Egorov, K.I. Rybakov, V.E. Semenov, Y.V. Bykov, O.N. Kanygina, E.B. Kulumbaev, V.M. Lelevkin, J. Mater. Sci. **42**, 6 (2007) 2097.
[12] Software plugins FracLac 2.5, http://rsb.info.nih.gov/ij/plugins /fraclac/FLHelp/Introduction.htm (ac. 20.04.2016).
[13] G.V. Vstovsky, *Elements of information physics*, MGIU, Moscow (2002) 260.
[14] B.B. Mandelbrot, *The geometry of nature: Russian translation*, Inst. Comp. Sci., Moscow (2002) 656.
[15] O.N. Kanygina, A.G. Chetverikova, M.M. Filyak, A.A. Ogerchuk, Glass Ceram. **72**, 11 (2016) 444.
[16] N.M. Astaf'eva, Physics-Uspekhi **39**, 11 (1996) 1085.
[17] K.I. Rybakov, E.A. Olevsky, V.E. Semenov, Scr. Mater. **66**, 12 (2012) 1049.
[18] V.I. Roldughin, Russian Chem. Rev. **72**, 11 (2003) 913.